\documentclass[journal,1p]{elsarticle}
\usepackage{amsmath}
\usepackage{amssymb}
\usepackage{amsfonts}
\usepackage{graphicx}
\usepackage{fullpage}
\usepackage{graphics}
\usepackage{multirow}
\usepackage{epsfig}
\usepackage{epstopdf}

\usepackage{newfloat}
\usepackage{subcaption}
\usepackage{gensymb}
\usepackage[table]{xcolor}
\usepackage{ulem}
\usepackage{varioref}
\usepackage{hyperref}
\usepackage{bm}

\usepackage{lineno,hyperref}
\modulolinenumbers[5]

\newcolumntype{H}{>{\setbox0=\hbox\bgroup}c<{\egroup}@{}} 

\graphicspath{{./}{./arxiv_images/}}

\journal{Journal of \LaTeX\ Templates}

\begin{document}

\begin{frontmatter}
\title{Impact of Black Swan Events on Ethereum Blockchain ERC20 Token Transaction Networks}

\author[mymainaddress1]{Moturi Pradeep}
\author[mymainaddress2]{Uday Kumar Reddy Dyapa}
\author[mymainaddress3]{Sarika Jalan}
\ead{sarika@iiti.ac.in}
\author[mymainaddress2]{Priodyuti Pradhan \corref{mycorrespondingauthor}}
\ead{priodyutipradhan@gmail.com}
\address[mymainaddress1]{Department of Physics \& Astrophysics, University of Delhi, India.}
\address[mymainaddress2]{Department of Computer Science \& Engineering, Indian Institute of Information Technology Raichur, Karnataka - 584135, India}
\address[mymainaddress3]{Complex Systems Lab, Department of Physics, Indian Institute of Technology Indore, Khandwa Road, Simrol, Indore-453552, India}

\begin{abstract}
The Ethereum blockchain and its ERC20 token standard have revolutionized the landscape of digital assets and decentralized applications. ERC20 tokens developed on the Ethereum blockchain have gained significant attention since their introduction. They are programmable and interoperable tokens, enabling various applications and token economies. Transaction graphs, representing the flow of the value between wallets within the Ethereum network, have played a crucial role in understanding the system's dynamics, such as token transfers and the behavior of traders. Here, we explore the evolution of daily transaction graphs of ERC20 token transactions, which sheds light on the trader's behavior during the Black Swan Events -- $2018$ crypto crash and the COVID-19 pandemic. By using the tools from network science and differential geometry, we analyze $0.98$ billion of ERC20 token transaction data from November $2015$ to January $2023$. Our analysis reveals that ERC20 financial ecosystem has evolved from a localized wealth formation period to a more mature financial ecosystem where wealth has dispersed among the traders in the network after the crypto crash and during the pandemic period. Before the crash, most sellers only sell the tokens, and buyers only buy the tokens. However, after the crash and during the pandemic period, sellers and buyers both performed buying and selling activities. In addition, we observe no significant negative impact of the COVID-19 pandemic on user behavior in the financial ecosystem. 
\end{abstract}

\begin{keyword}
\texttt{Ethereum Blockchain \sep Financial Networks \sep  Transaction graph \sep Forman-Ricci Curvature}
\end{keyword}

\end{frontmatter}


\section{Introduction}
The growing enthusiasm worldwide to understand the financial ecosystem is largely due to several Black Swan events like the credit crisis of $1772$, the great depression of $1929-39$, the OPEC oil price shock of $1973$, the Asian crisis of 1997, and $2007-2008$ financial crisis \cite{black_swan_events_2008, financial_crisis}. Modeling a financial system as a network has helped us to understand a wide range of phenomena crucial for financial professionals, economists, and researchers \cite{finantial_networks_nat_phys_2021}. Analysis of a financial network sheds light on underlying salient features which may not be evident without the holistic approach of network science \cite{Traditional Networks}, thereby providing a  better understanding of how the traders interact with each other and how their interactions affect the whole system \cite{Systemic Risk, contagion}. However, it was not as much of a success as thought it would be, as there exist constraints in modeling the underlying networks of traditional financial systems arising due to many reasons; for example,  confidentiality issues where a financial institution or bank may not fully provide all the transaction details due to the intellectual property restrictions, and privacy rules. Consequently,  various features of the traditional economy still need to be explored. We consider the Ethereum Blockchain transaction data to analyze the trader's behavior during the $2018$ crypto crash and the COVID-19 pandemic \cite{Ethereum White Paper, Ethereum}.  

Ethereum blockchain may be modeled using networks as entities are connected through transactions of many assets \cite{temporal_analysis_blockchain, transaction_network_2018, transaction_network_2020}. Initially, transaction graphs within the ERC20 token financial ecosystem were relatively simple and characterized by straightforward transfers between token holders. However, as the financial ecosystem evolved, transaction graphs became increasingly complex, reflecting the growth and diversification of token-related activities \cite{erc20_token_networks_2019}. New patterns, including token swaps  \cite{Uniswap}, lending protocols \cite{Decentralized Finance}, and decentralized exchanges  \cite{Decentralized_Exchange}, led to intricate and intertwined transaction graphs. Analyzing and understanding ERC20 transaction graphs has become crucial for researchers, developers, and regulators seeking to comprehend token movements, identify patterns, and assess network health. Tools and techniques, such as graph analysis algorithms and visualization frameworks, have emerged to extract meaningful insights from transaction graphs, aiding in risk assessment, fraud detection, and market analysis within the Ethereum Financial Ecosystem \cite{financial_systems, blockchain_risk_analysis_2020, Evolutionary Dynamics, Phishing Networks, Ethereum Structure}. A few existing works analyze the ERC20 transaction data and crypto crash \cite{crypto_crash_complex_systems}. However, the impact of critical events and the behavior of traders still needs to be discovered as the system is continuously evolving.

This article studies the impact of the crypto crash and the COVID-19 pandemic on the trader's behavior of the ERC20 token transactions. We use network methods to model and analyze the structural and dynamic behavior of the blockchain's transaction graphs. To examine the financial ecosystem, we create daily transaction graphs from November $2015$ to January $2023$. We investigate the evolution of traders' behavior in the Ethereum blockchain. Our analysis unveils that before the crash, most sellers only sell the tokens, and buyers only buy the tokens. Few transactions among the small traders lead to the localization of wealth among the individual traders. However, after the crash and during the pandemic, the seller sells the token, and buyers buy the token. But at the same time, the seller buys the tokens, and the buyer sells the token leading to the dispersal of the wealth among the traders and making the ERC20 financial ecosystem more stable during the pandemic. In addition, we show no significant negative effect of the COVID-19 pandemic on user behavior in the financial system. 

The article is organized as follows: Section 2 discusses preliminaries of the Ethereum Blockchain and ERC20 tokens. Section 3 illustrates the details of the extraction and preprocessing of the ERC20 transaction data and the modeling of the daily transaction network. It also contains the notations and definitions used in the later discussions. Section 4 explains the results and analysis. Finally, section 5 summarizes the current study and discusses the open problems for further investigation. 

\section{Preliminary}
Blockchain is an underlying technology on which the famous cryptocurrency, BitCoin, was built; nowadays, blockchain applications are widespread, which cover supply chains, financial services, healthcare, and public registers \cite{blockchain, block_chain_supply_chain,blockchain_2016, blockchain_2020}. The core components of blockchain are transparency and trustlessness, through which transactions are validated and broadcasted. In the blockchain financial ecosystem, a block comprises several transactions and is linked to its previous block via a digital link, thus forming a chain of blocks. 

\subsection{Ethereum Blockchain}
In the year 2015, Ethereum came into existence \cite{Ethereum White Paper}. Ethereum allows for the creation and direct peer-to-peer exchange of digital assets without intermediaries. Ethereum platform is a software built on blockchain technology that enables the creation of cryptocurrency (Ether), crypto-assets (e.g., ERC20 tokens, ERC721 tokens \cite{NFT's}), and Decentralized Applications (DApps) \cite{decentralized application}. The Ethereum blockchain is a digital ledger where Ether and crypto-assets can be securely stored and exchanged. Ether is the backbone of the platform, which facilitates transactions and pays for the deployment of smart contracts on the Ethereum Blockchain.

The primary focus of the platform is to use decentralized blockchain technology for smart contracts \cite{blockchain_2020, Smart Contracts}. The smart contract is a computer protocol used to create and develop DApps, and crypto assets. Smart contracts are conditional codes on the blockchain executed when smart contract conditions are met. In other words, they are $``if \ldots then \ldots"$ statements written in the form of code and deployed on the blockchain. For example, a certificate contract in which the smart contract will provide the certificate when the participants attend the required number of classes of a course and score more than or equal to 60 marks in that course. The usage of smart contracts is very diverse and includes digital identity, real estate \cite{Real Estate}, insurance, flash loans \cite{Decentralized Finance}, gaming \cite{Gaming}, and decentralized finance \cite{Decentralized Finance}. 

\begin{figure}[tbh]
\begin{center}
\includegraphics[width=5.6in, height=4in]{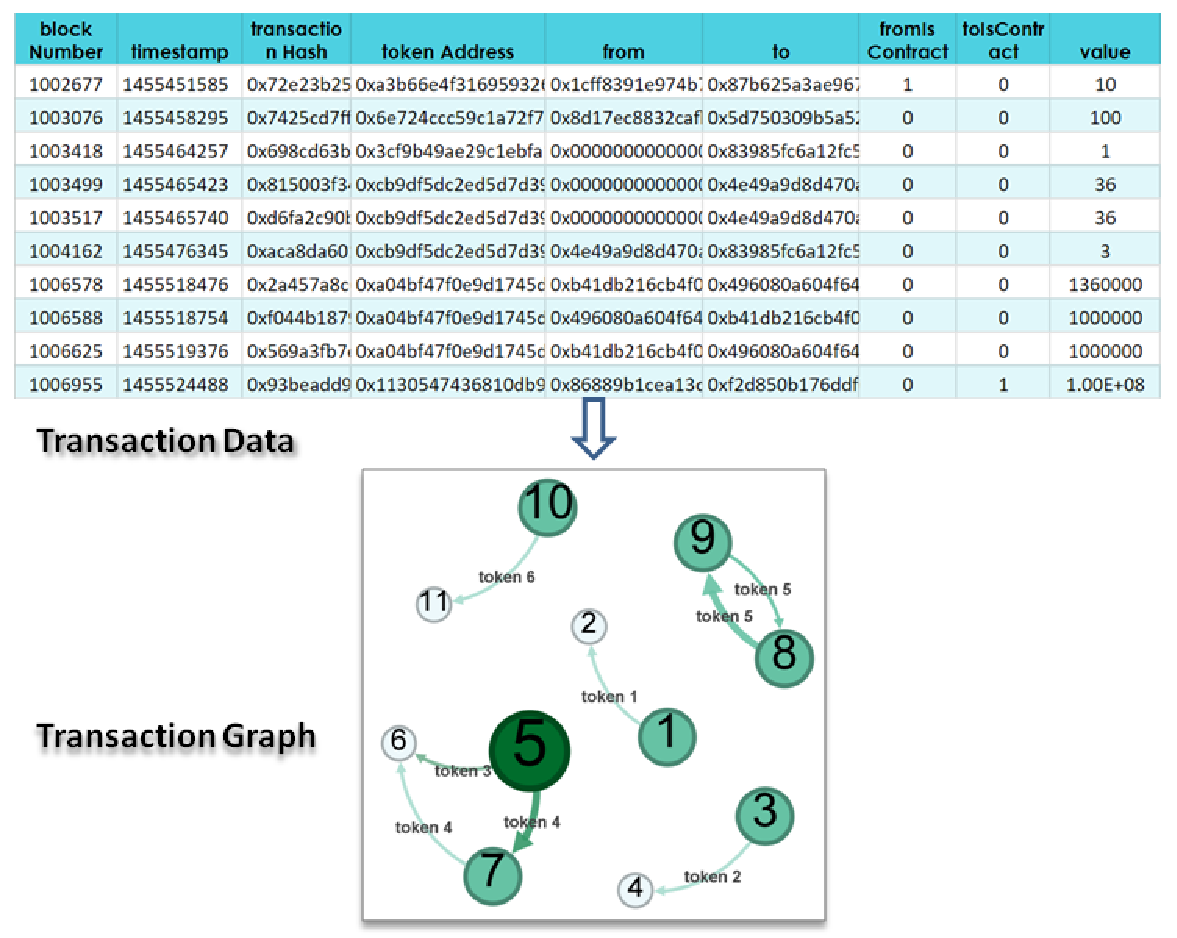}
\caption{Illustrate the ERC20 token transaction data over the Ethereum blockchain and the associated transaction graph. For simplicity, we assign a unique integer number corresponding to the `from', `to', and `token Address' columns. Here, `from' is the seller's, `to' is the buyer's wallet, and the edge label shows which tokens are traded. The edge thickness represents multiple transactions of the same tokens between buyer-seller. For instance, between node $5$ and $7$, two transactions of `token 4'.}
\label{sample_table}
\end{center}
\end{figure}
In the Ethereum Financial ecosystem, users interact with the Ethereum network through their Ethereum account. With the help of accounts, users can transfer assets, create or invoke smart contracts, and interact with DApps \cite{Ethereum White Paper}. A user account consists of a $40$-byte public address (like a bank account number) with the prefix $``0x"$ (e.g., $0x52d3fbd8fc248c\ldots25c37c5f5$), which other users use to transfer assets. A transaction in the Ethereum platform can execute various things, such as transferring assets (ERC20 tokens), deploying smart contracts, and triggering the smart contract \cite{Ethereum White Paper}. To deploy a smart contract, a person uses an Ethereum account and sends a transaction containing compiled code of the smart contract without the recipient of the transaction \cite{Deploy Contract}. This article limits our discussion to the transactions related to the ERC20 tokens.

\subsection{ERC20 Token}
Ethereum Blockchain platform provides a more accessible opportunity for companies and individuals to develop blockchain products instead of building their own blockchain platform \cite{erc20_token_networks_2019}. The Ethereum Request for Comment 20 (ERC20) standard allows developers to create smart contract-enabled tokens that can be used with other products and services, such as DApps on the Ethereum network, which started on Nov $2015$ \cite{Token Classification}. Sometimes we refer to tokens and coins as the same, but they are different in what they represent and their functions. In both cases, they are digital assets, but the coin is a native asset of the platform, which facilitates operations on the platform, whereas tokens are built on the platform for the creation and flow of wealth. For instance, Ether is the native coin of Ethereum, and Polygon MATIC \cite{matic} and USDT \cite{Tether} are tokens built on the Ethereum platform. 

In the Ethereum Blockchain, the digitalization of the value of a particular asset into tradeable digital units is known as tokenization, and the digital assets are represented as tokens. Tokens allow a seamless, borderless, and almost free flow of value in the form of digital assets across the globe. Once any product is tokenized, these tokens can be managed, detected, accounted for, and leveraged in the context of incentives that may promote fair wealth. For example, XAUt (Tether Gold) is a token representing gold as a digital token on the Ethereum platform. One XAUt token equals $31.1035$ grams of gold. Hence, XAUt tokens digitally represent the value associated with gold assets so that they can be traded across the globe using the Ethereum Platform. The above example of the XAUt token is an asset-backed token; there are various other types of tokens on the Ethereum platform with multiple functions and features \cite{Token Classification}. 
\begin{figure}[tbh]
\centering
\includegraphics[width=5.6in, height=2.2in]{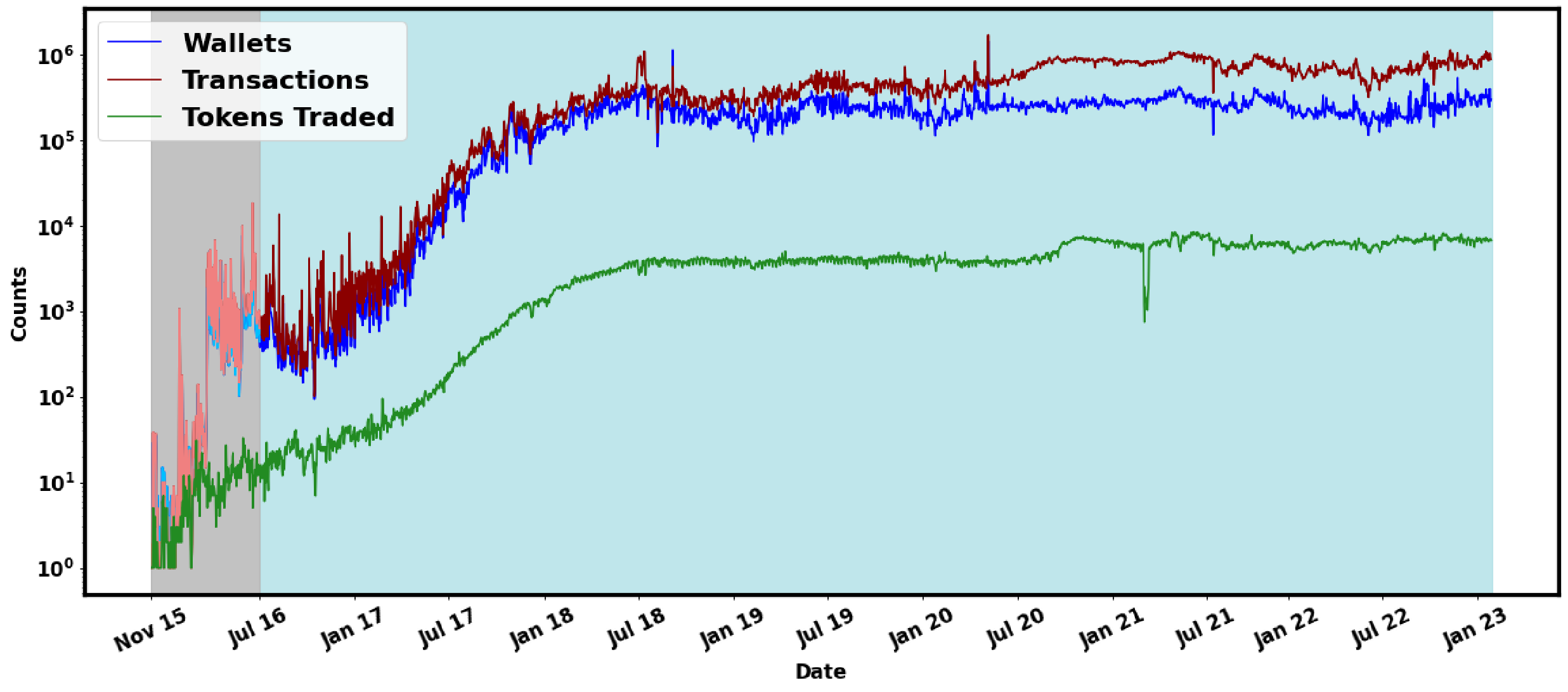}
\caption{Portray the evolution of Ethereum blockchain transaction data as wallets (nodes), transactions (edges), and the number of unique traded tokens. Tokens are the attributes on the edges. We examine the daily transaction graph from November $2015$ to January $2023$. The shaded region reflects the testing period of ERC20 tokens. We observe a rapid increase in all three variables between July $2016$ and July $2018$. After that, the number of nodes reaches stability, and edges gradually increase, showing the growing activity between the nodes.}
\label{Nodes_Edges}
\end{figure}
The ERC20 token can be created by any individual or organization that defines the rules governing them, such as monetary policy, token features, user incentive systems, etc. The current market cap of Ethereum is approximately $\$229.56B$, and ERC20 tokens are approximately $\$112.7B$, around $49\%$ of the total Ethereum blockchain \cite{market_cap_ref}. A high market capitalization implies that the market highly values the asset, and our interest lies in studying the trader's behavior involved in ERC20 token transaction.

\section{ERC20 Token Transaction Data and Network Modeling}
\subsection{Transaction Data sets}
To analyze the underlying network of ERC20 transactions in the Ethereum Blockchain, we use the past $8$ years of ERC20 transaction data \cite{ERC_datasets}. We analyzed $982,119,361$ ERC20 token transaction data from November $2015$ to January $2023$. The data set consists of $9$ columns (Fig. \ref{sample_table}); each column gives us specific information regarding the ERC20 transaction data and can be summarized as follows. 
\begin{enumerate}
    \item \textbf{blockNumber:} block number in which the transaction information has been stored.
    \item \textbf{timeStamp:} time in which the block was minted, and every transaction in a block has the same timestamp. 
    \item \textbf{transactionHash:} unique identifier that serves as proof of transaction validation.
    \item \textbf{tokenAddress:} the hash value refers to the actual smart contract address of the ERC20 token, which also acts as an identifier for an ERC20 token.
    \item \textbf{from:} address of the sender of ERC20 token
    \item \textbf{to:} address of the receiver of ERC20 token
    \item \textbf{fromisContract:} if this field value is $1$, it signifies the `from' column is a smart contract address otherwise an externally owned account address.
    \item \textbf{toisContract:} if this field value is $1$, it signifies the `to' column is a smart contract address otherwise an externally owned account address.
    \item \textbf{value:} tells about the number of tokens transferred
\end{enumerate}
Each row provides information about an ERC20 token transaction in the data set. The `from' and `to' columns are the addresses between whom the transaction has taken place (Fig. \ref{sample_table}). For our analysis, we use four columns `timeStamp', `tokenAddress', `from', and `to'. The `timestamp' column is in seconds, which we convert into ($YYYY-MM-DD$) format. For instance, after transforming the timestamp in Fig. \ref{sample_table}, $1455451585$ becomes $2016-02-14$ where base time ($1970-01-01$) is considered standard time $00:00:00$ UTC \cite{time_format}. The rest of the three columns' data are in hash value which is very difficult to analyze. For better viewing and analyzing the data, we iterated over the `from' and `to' columns and mapped every unique address with a unique integer number. The same iteration process is carried out for the `tokenAddress' column. Finally, we divide the whole data set in day-wise.

\begin{figure}[tbh]
\centering
\includegraphics[width=5.6in, height=2.2in]{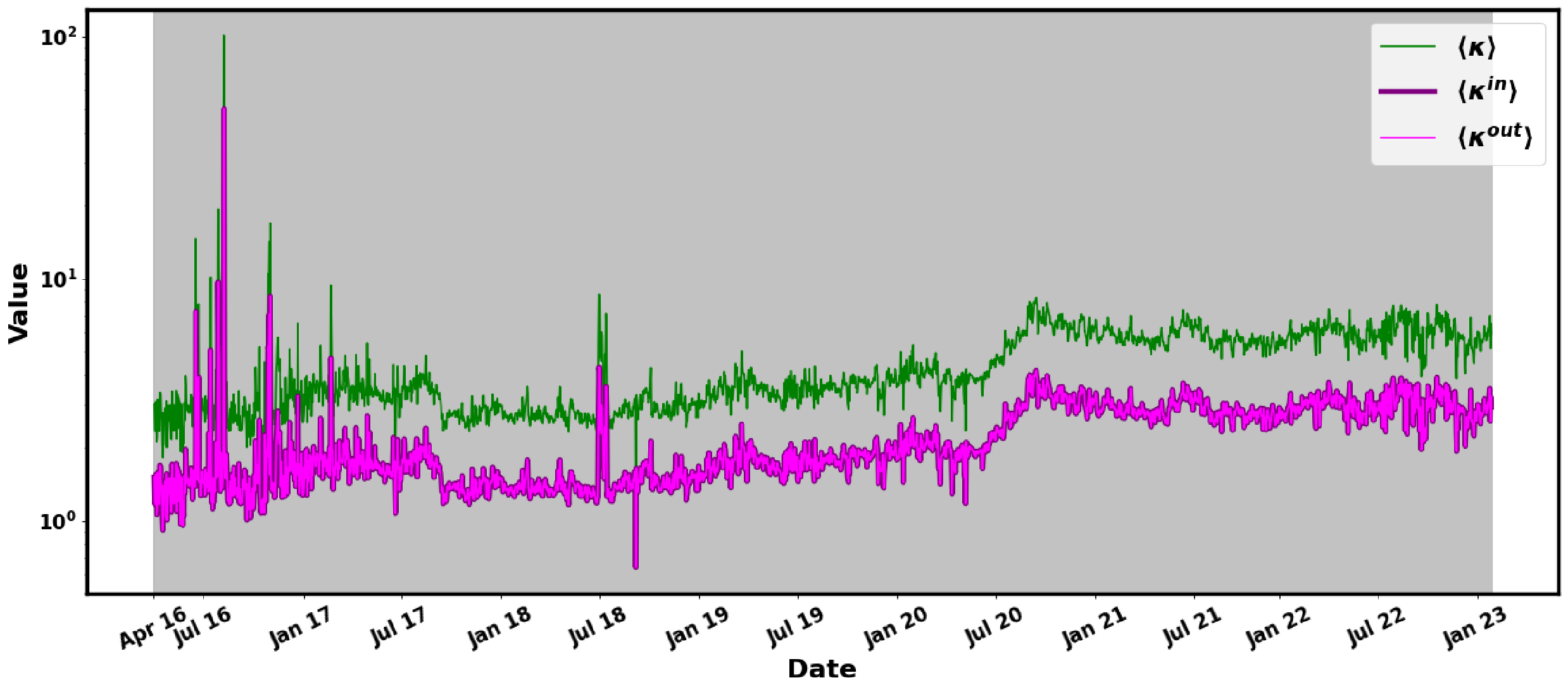}
\caption{Dynamic behavior of average degree ($\langle k \rangle$), in-degree ($\langle k^{in} \rangle$), and out-degree ($\langle k^{out} \rangle$) of daily transaction graphs. The average degree of the transaction graph provides the average number of transactions a wallet carries out in a day. The average in-degree (out-degree) is around $3$. The fluctuations in the inception period arise due to a large number of parallel edges (transactions) between a pair of nodes (wallets) during the testing of the ERC20 token.}
\label{avg_degree}
\end{figure}

\subsection{Transaction Network}
To model the ERC20 transaction data, we use the graph model \cite{complex_networks_2001}. In the Ethereum Financial Ecosystem, wallets are the nodes that buy or sell ERC20 tokens, and transactions between two wallets are the edges (links). For instance, let wallets $A$ make a transaction in which $A$ sends $1$ token to $B$, then the link will be directed from $A$ to $B$ ($A \rightsquigarrow B$). Further, if a wallet, $A$ makes $2$ transactions with $2$ other wallets ($B$ and $C$) in a day, then there will be two directed edges between the nodes as $A \rightsquigarrow B$ and $A \rightsquigarrow C$. Here, the node $A$ has out-degree $2$ and $B, C$ both having in-degree $1$. A node can have $10$ edges with another node if it makes $10$ transactions with the same node in a day with different tokens; then, there will be $10$ parallel edges between them. Therefore, ERC20 token transaction graph is a multi-edges directed graph consisting of source and target nodes, where source nodes are the wallets that sell the ERC20 tokens, and target nodes are the wallets that buy the ERC20 tokens. We can think of tokens as the attributes on the edges of the transaction graphs (Fig. \ref{sample_table}). The transaction graph for a given day $t$, represented as $\mathcal{G}_t$($V_t$, $E_t$) where set of vertices ($V_t$) consists of all wallets trading during that day as \cite{financial_systems}
\begin{equation}
V_t = \{\; v \;||\; \text{wallets}\; v \;\text{buy or sell any assets at day}\; t \}  
\end{equation}
and the set of edges $E_t \subseteq V_t \times V_t$ is defined as:
\begin{equation} 
E_t = \{\; (u,v)\; ||\; \text{wallet $u$ sell to wallet}\; v \;\text{any asset at day}\; t \}
\end{equation}
We denote the adjacency matrices corresponding to multi-edge directed graph $\mathcal{G}_t$ as ${\bf A}_t \in \mathbb{R}^{n_t \times n_t}$ and which can be defined as $a_{ij}=l$ if there are $l$ edges from $i$ to $j$ and $0$ otherwise. The out-degree of a node, $i$ on day $t$ can be represented as $k_{i,t}^{out}=\sum_{j=1}^{n_t} a_{ij}$ and in-degree as $k_{i,t}^{in}=\sum_{j=1}^{n_t} a_{ji}$. The average out-degree and in-degree of $\mathcal{G}_t$ can be defined as  $\langle k_{t}^{out} \rangle = \frac{1}{n_t}\sum_{i=1}^{n_t} k_{i,t}^{out}$ and $\langle k_{t}^{in} \rangle = \frac{1}{n_t}\sum_{i=1}^{n_t} k_{i,t}^{in}$, respectively. Here, we consider number of wallets participated on day $t$ as $|V_t|=n_t$, and number of transactions as  $|E_t|=\sum_{i=1}^{n_t} k_{i,t}^{out}=\sum_{i=1}^{n_t} k_{i,t}^{in}=m_t$, thus $\langle k_{t}^{out} \rangle=\langle k_{t}^{in} \rangle$. Further, a node that sent the maximum number of transactions in a day as a max-out-degree node and represented as $k_{max,t}^{out}=\operatorname*{max}_{i\in V_{t}} k_{i,t}^{out}$. Similarly, a node that receives a maximum number of transactions in a day as max-in-degree and defined as $k_{max,t}^{in}=\operatorname*{max}_{i\in V_{t}} k_{i,t}^{in}$. We can define sets containing all the nodes having out-degree equal to $\alpha$ as $\mathcal{D}_{\alpha,t}^{out}=\{i\in V_t||\;k_{i,t}^{out}=\alpha, \alpha=1,2,\ldots,k_{max,t}^{out}\}$ and in-degree equal to $\beta$ as $\mathcal{D}_{\beta,t}^{in}=\{i\in V_t||\;k_{i,t}^{in}=\beta,\beta=1,2,\ldots,k_{max,t}^{in}\}$, where  $N_{\alpha,t}^{out}=|\mathcal{D}_{\alpha,t}^{out}|$ and $N_{\beta,t}^{in}=|\mathcal{D}_{\beta,t}^{in}|$, are the number of elements inside the sets \cite{financial_systems}. Hence, sets containing all the nodes having out-degree and in-degree equal to $1$ as $\mathcal{D}_{1,t}^{out}=\{i\in V_t||\;k_{i,t}^{out}=1\}$ and $\mathcal{D}_{1,t}^{in}=\{i\in V_t||\;k_{i,t}^{in}=1\}$, where $N_{1,t}^{in}=|\mathcal{D}_{1,t}^{in}|$, and $N_{1,t}^{out}=|\mathcal{D}_{1,t}^{out}|$. From the economic perspective -- $k^{out}_{max,t}$ is a wallet that is a maximum selling hub, $k^{in}_{max,t}$ is a wallet that is a maximum buying hub, $N^{out}_{1,t}$ is the number of wallets which sell once and  $N^{in}_{1,t}$ is the number of wallets which buy once on a daily basis. Note that in the later discussion, we omit $t$ from the above notations for convenience.

\begin{figure}[tbh]
    \centering
    \includegraphics[width=6in, height=2.5in]{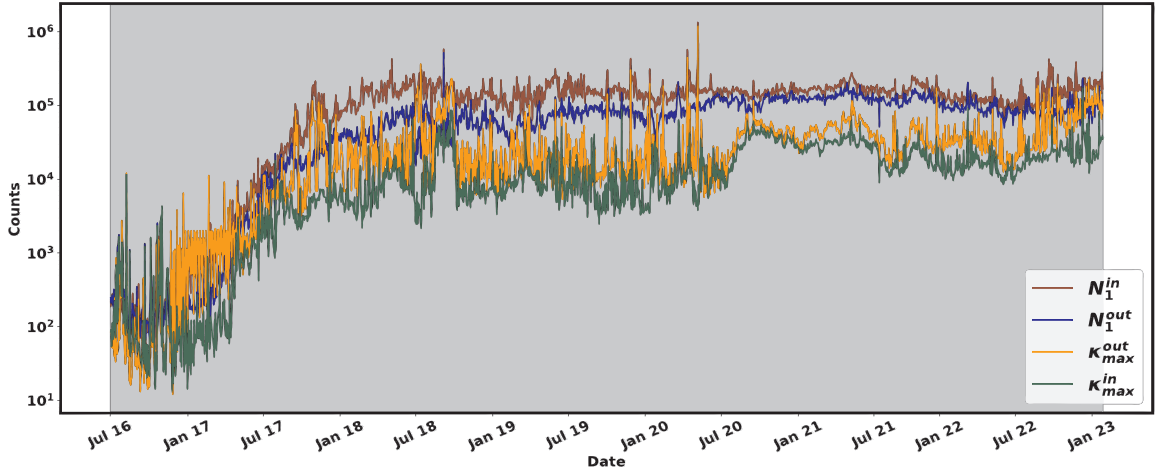}
    \caption{Represents the dynamics of the maximum selling hub ($k_{max}^{out}$), maximum buying hub ($k_{max}^{in}$), number of wallets buying once ($N^{in}_{1}$), and number of wallets selling once ($N^{out}_{1}$). We observe that $N^{in}_{1}$ has more steeper increase than $N^{out}_{1}$ until July $2018$, after that $N^{in}_{1}$ reaches its stability whereas there is gradual increase in $N^{out}_{1}$. After July $2020$, we observe strong co-movement between $N^{in}_{1}$ and $N^{out}_{1}$.}
   \label{completedynamics}
\end{figure}

\section{Results and Discussion}
In January $2018$, the Ether price reached its record high of $\$1431$, and by the middle of December 2018, the Ether price was down by $94\%$ \cite{Crypto Crash}. This period was marked as the $2018$ Crypto Market Crash, where various other cryptocurrencies also hit record lows \cite{Cryptocurrency Buble}. On the other hand, in March $2020$, COVID-19 was declared a pandemic by the World Health Organization, resulting in severe societal and economic ramifications worldwide \cite{Economy Effect}. During these events, significant changes occurred in the trading behavior of the Ethereum ERC20 Financial Ecosystem. To understand, we analyze the behavior of the daily transaction graphs. 

\subsection{Dynamics of the System}
After the inception of Ethereum ERC20 tokens, the number of wallets and transactions was lower; however, after July 2016, we can see a notable increase in the development of nodes, edges, and the volume of tokens traded over time (Fig. \ref{Nodes_Edges}). After July $2018$, the everyday number of wallets (nodes) involved in trading is approximately constant. Still, the number of daily transactions (edges) increases gradually, which infers the growing activity between the wallets of the Ethereum ERC20 Financial Ecosystem. From the daily transaction graph, we can also predict that on average, $10^5$ wallets perform around $10^5$ transactions, and on average, $10^3$ distinct types of tokens traded (Fig. \ref{Nodes_Edges}). 

Additionally, for the initial period, the average number of transactions carried out by wallets per day is around $4$ and gradually grows to around $6$ after July $2020$ (Fig. \ref{avg_degree}). However, if we separately look into the average out-degree and in-degree, it is close to $3$. On the contrary, Fig. \ref{completedynamics} reveals that the max-out-degree ($k_{max}^{out}$) is very large as compared to the average out-degree ($\langle k^{out} \rangle$). Also, we can notice a large number of nodes having one out-degree ($N_{1}^{out}$). Similar, behavior for the max-in-degrees ($k_{max}^{in}$, $\langle k^{in} \rangle$ and $N_{1}^{in}$). It infers degree distribution might be heavy-tailed where $N_{1}$ and $k_{max}$ are the extreme points of the degree distribution \cite{financial_systems}. If we randomly pick a daily transaction graph, it shows a heavy-tailed degree distribution for both the out-degrees and in-degrees. The out-degree distribution is of the seller's wallet, and the in-degree distribution is of the buyer's wallet of the ERC20 token. The distribution clearly shows that the Ethereum ERC20 Financial Ecosystem follows heavy-tailed distribution for daily transaction graphs, which coincides with numerous previous works showing that the degree distribution of blockchain transaction data is heavy-tailed \cite{erc20_token_networks_2019, financial_systems}. 
 
\begin{figure}[ht]
    \centering
\includegraphics[width=6in, height=2.2in]{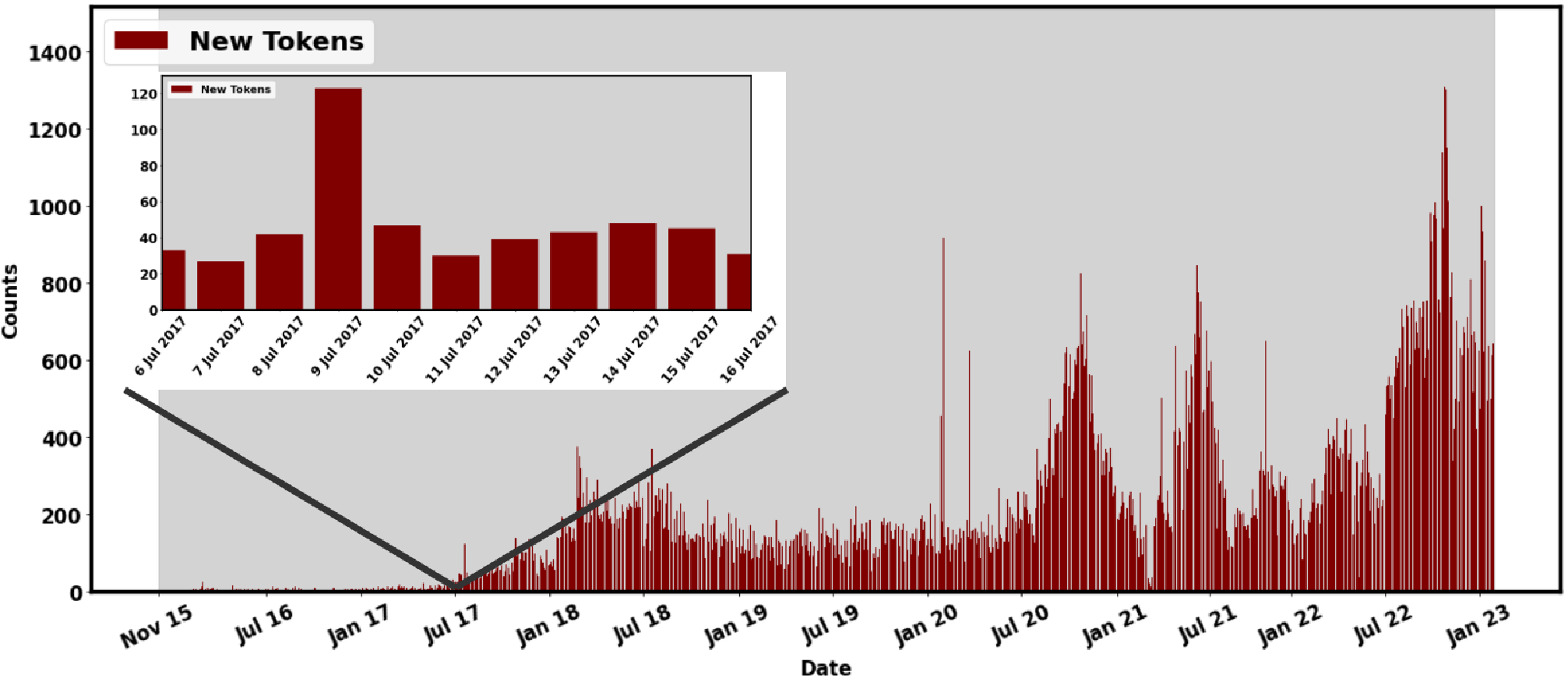}
\caption{Evolution of Tokens. Presents the dynamics of the addition of new ERC20 tokens to the network. For each day, we extract the count of new tokens added to the network. For instance, on $7^{th}$ July $2017$, $25$ new tokens are added, $8^{th}$ July $2017$, $40$ new tokens are added to the network, and so on. We observe that until July $2018$, there is an increase in the addition of new tokens to the network, but after that, the count remains approximately constant until July $2020$. There exists a volatile behavior of the token evolution during the COVID-19 pandemic. The total number of unique tokens traded over the whole period is $301428$.}
\label{token_evolution}
\end{figure}

To get insights on the buyers' and sellers' behavior before and after the crypto crash, we examine the dynamical behavior of the extreme points of degree distribution -- maximum selling hub ($k^{out}_{max}$), number of wallets which sell once ($N^{out}_{1}$), maximum buying hub ($k^{in}_{max}$), and the number of wallets which buy once ($N^{in}_{1}$) daily (Fig. \ref{completedynamics}). We observe that until July $2018$, all four variables grow substantially. But after that, the number of wallets buying once daily reached stability. In contrast, the number of wallets selling once is still gradually increasing but not substantially, and there is a decrease in the maximum selling and buying hub until July $2020$ (Fig. \ref{completedynamics}). 

\begin{figure}[ht]
    \centering
\includegraphics[width=6in, height=2in]{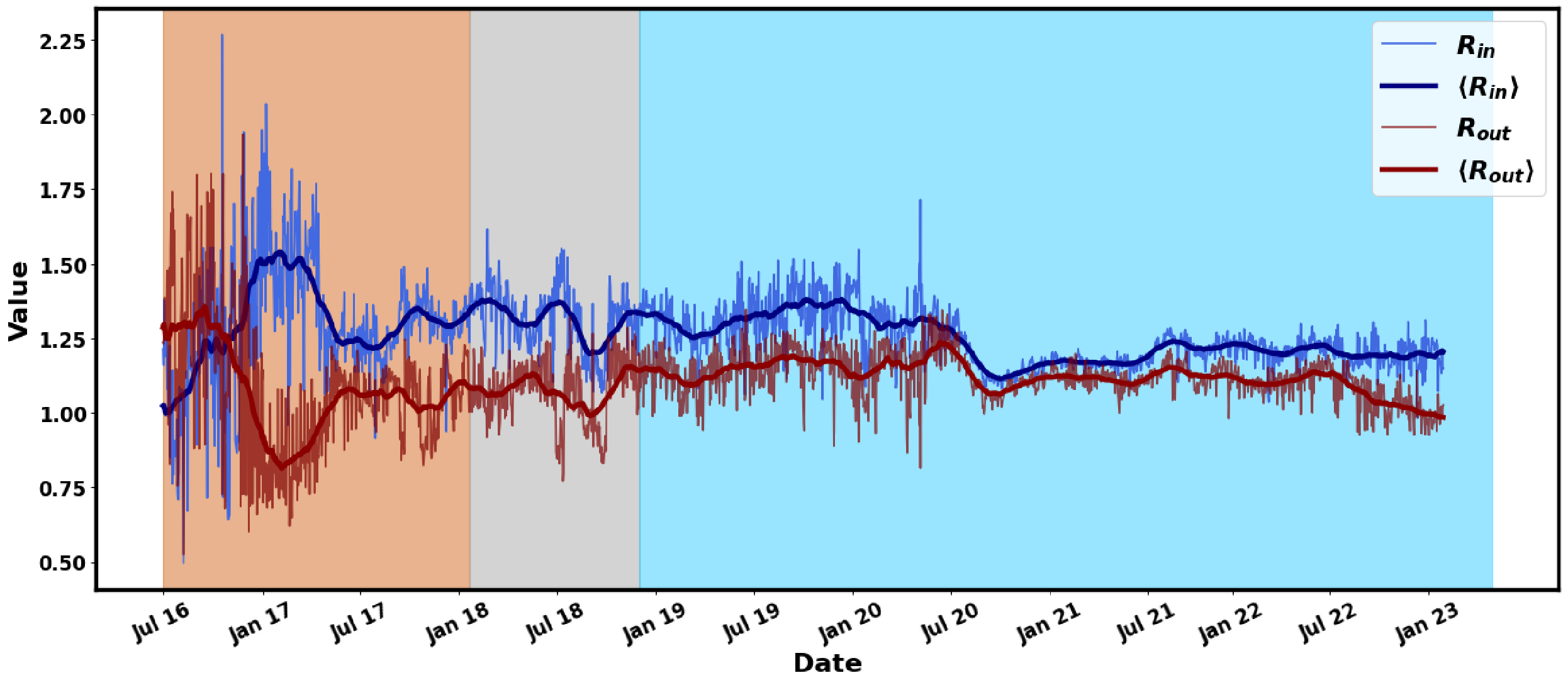}
\caption{Dynamical behavior of buyers and sellers ratios is represented as in-degree ratio ($R_{in}$) and out-degree ratio ($R_{out}$), respectively. For each transaction graph, we calculate $R_{in}$, $R_{out}$ using Eq. (\ref{Rin}). To observe the evolution of the Financial Ecosystem's dynamics, we calculate the moving average of $R_{in}$ and $R_{out}$ ($\langle R_{in} \rangle$ and $\langle R_{out} \rangle$) for each day. From July 2016 to July 2018, we observe an anti-phase oscillation between $R_{in}$ and  $R_{out}$. However, after July $2018$, we see a change in the dynamics of $R_{in}$ and $R_{out}$ with co-movement between the two, which grows stronger after July $2020$ (COVID-19 period). For a given day $t$, $\langle R_{in} \rangle$ is calculated by taking the mean of window length $p+t+s$ that includes the $R_{in}$ value of the day $t$, $p$ is the number of $R_{in}$ values preceding the day $t$ and $s$ is the number of $R_{in}$ values succeeding the day $t$. The window length is truncated at the initial and final days when there are insufficient $R_{in}$ values to fill the window. The mean value is taken over only the $R_{in}$ that fill the window. Here, we consider the window size to be $70$. For the initial days, the size of $p$ is dynamically growing, and $s$ is kept constant until $p$ equals $34$. For final days, the size of $s$ is dynamically growing, and $p$ is kept constant when the successive days are less than $35$. Similarly, we calculate the $\langle R_{out} \rangle$ values over time.}
\label{Rin_Rout}
\end{figure}

Furthermore, we calculate the ratios between the extreme points of the degree distribution for each day. In that case, we can observe significant changes in the network's global dynamics during the $2018$ crypto crash and the COVID-19 pandemic. We can define the ratio as follows \cite{financial_systems}
\begin{equation}\label{Rin}
R_{in}(\mathcal{G}_t)=\frac{\log(N_{1,t}^{in})}{\log(k_{max,t}^{in})}, \;\;\text{and}\;\;
R_{out}(\mathcal{G}_t)=\frac{\log(N_{1,t}^{out})}{\log(k_{max,t}^{out})}
\end{equation}
The ratios show the interplay between the buyers' and sellers' behavior of the Ethereum ERC20 token transactions and provide insight into their evolution over time. We can observe high volatility in the dynamics of the ratios (Fig. \ref{Rin_Rout}).
However, close observation of $R_{in}$ and $R_{out}$ reveals a change in the dynamical behavior of the ratios before and after the crypto crash, which suggests a change in the trader's trading behavior. The moving average of the ratios denoted as $\langle R_{in} \rangle$ and $\langle R_{out} \rangle$ can prominently show the behavioral changes of the buyers and sellers. We remark that before July $2018$, 
\begin{figure}[tbh]
    \centering
\includegraphics[width=4.2in, height=4in]{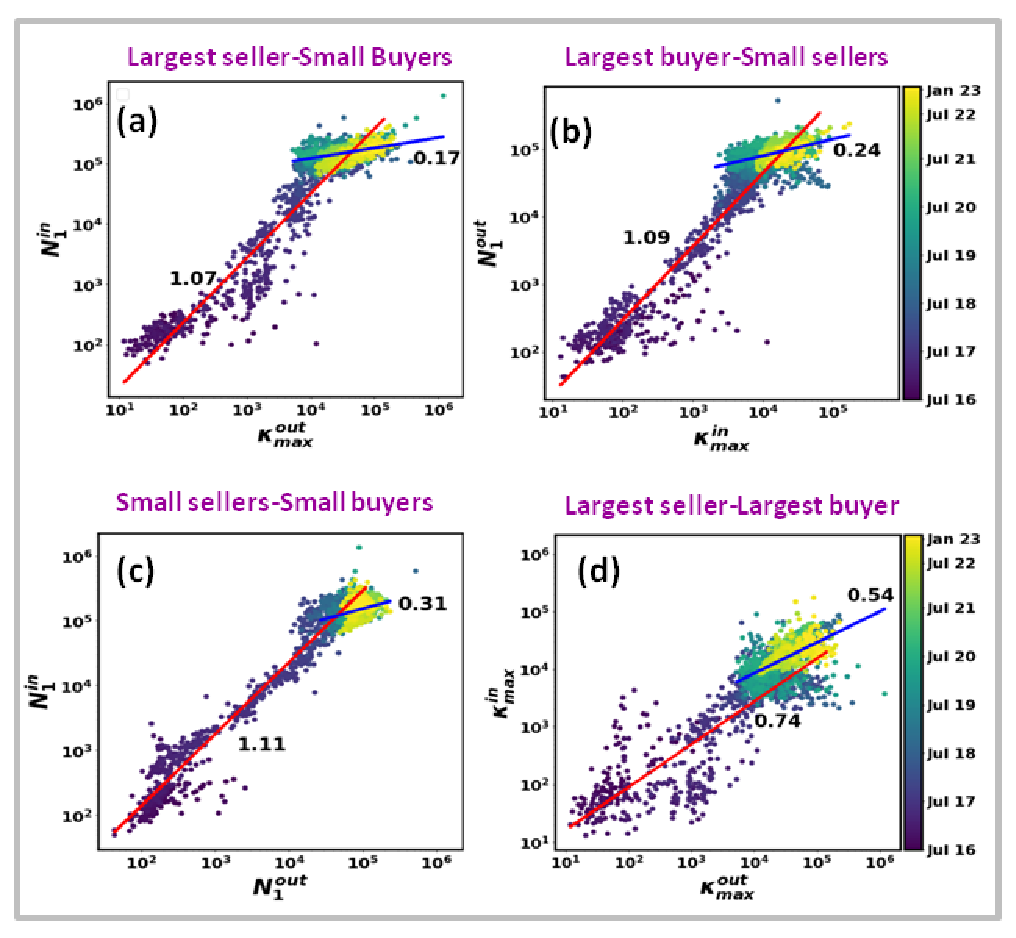}
    \caption{Buyer's and seller's behavior. Illustrate the relation between (a) largest seller ($k^{out}_{max}$) vs. small buyers ($N^{in}_{1}$) (b) largest buyer ($k^{in}_{max}$) vs. small seller ($N^{out}_{1}$), (c) small sellers vs. small buyers and (d) largest seller vs. largest buyer. The color bar corresponds to the date. We calculate the slope between the entities for two different periods. The red line refers to the slope from July $2016$ to July $2018$, and the blue line refers to the slope from July $2018$ to January $2023$. We can observe a large slope value in the initial period for panels (a-c) ($k^{out}_{max}$ vs. $N^{in}_{1}$, $k^{in}_{max}$ vs. $N^{out}_{1}$ and $N^{out}_{1}$ vs $N^{in}_{1}$) and a decrease in the later periods. However, from July $2018$ to January $2023$, we can observe a larger slope value between $N^{out}_{1}$ vs $N^{in}_{1}$, and $k^{out}_{max}$ vs $k^{in}_{max}$ as compared to other panels.}
\label{corr}
\end{figure}
when buyers' activity ($\langle R_{in} \rangle$) increasing, sellers' activity ($\langle R_{out} \rangle$) decreasing and vice-versa (Fig. \ref{Rin_Rout}). We characterize this day-wise phenomenon in transaction graphs as anti-phased oscillations \cite{financial_systems}. Notably, after July $2018$, there was a co-movement of the buyers' and sellers' activity (Fig. \ref{Rin_Rout}). The daily transaction graph size is very large and dynamic, so it is difficult to understand the internal behavior. Therefore, we use the correlation measure and regression analysis among the variables in Eq. (\ref{Rin}). Anti-phase oscillation of $\langle R_{in} \rangle$ and $\langle R_{out} \rangle$ to each other in the initial period is resulted due to a strong correlation between entities in Eq. (\ref{Rin}) --  maximum selling hub ($k^{out}_{max}$) vs. number of wallets buying once ($N^{in}_1$), and maximum buying hub ($k^{in}_{max}$) vs. the number of wallets selling once ($N^{out}_1$) (Fig. \ref{corr}(a-b)). Simultaneously, the correlation between the number of wallets selling once vs. the number of wallets buying once, and a weak correlation between the maximum buying hub and maximum selling hub (Fig. \ref{corr}(c-d)). The slopes in the regression analysis also show that after July $2018$, the value of the slope decreases (Fig. \ref{corr})(a-b). From the correlation and slope analysis, we might conclude that during the initial period, most of the transactions of the small traders are with big traders, fewer among small traders, and similarly, fewer transactions between big traders. 

However, after July $2018$, both $\langle R_{in} \rangle$ and $\langle R_{out} \rangle$ show co-movement to each other, which grows stronger over the period, especially after July $2020$ (COVID-19 period). We observe the co-movement of the ratios lead to a decrease in the correlations between the maximum selling hub and the number of wallets buying once (Fig.\ref{corr}(a)), as well as the maximum buying hub and the number of wallets selling once (Fig. \ref{corr}(b)). Simultaneously, there is an increase in the correlation between the number of wallets selling and buying once and between maximum buying and selling hubs (Fig.\ref{corr}(c-d)). One can notice the decrement of the slopes during the co-movement for $2$ relations and an increase in other $2$ relations. 
\begin{figure*}[tbh]
\centering
\includegraphics[width=6in, height=4in]{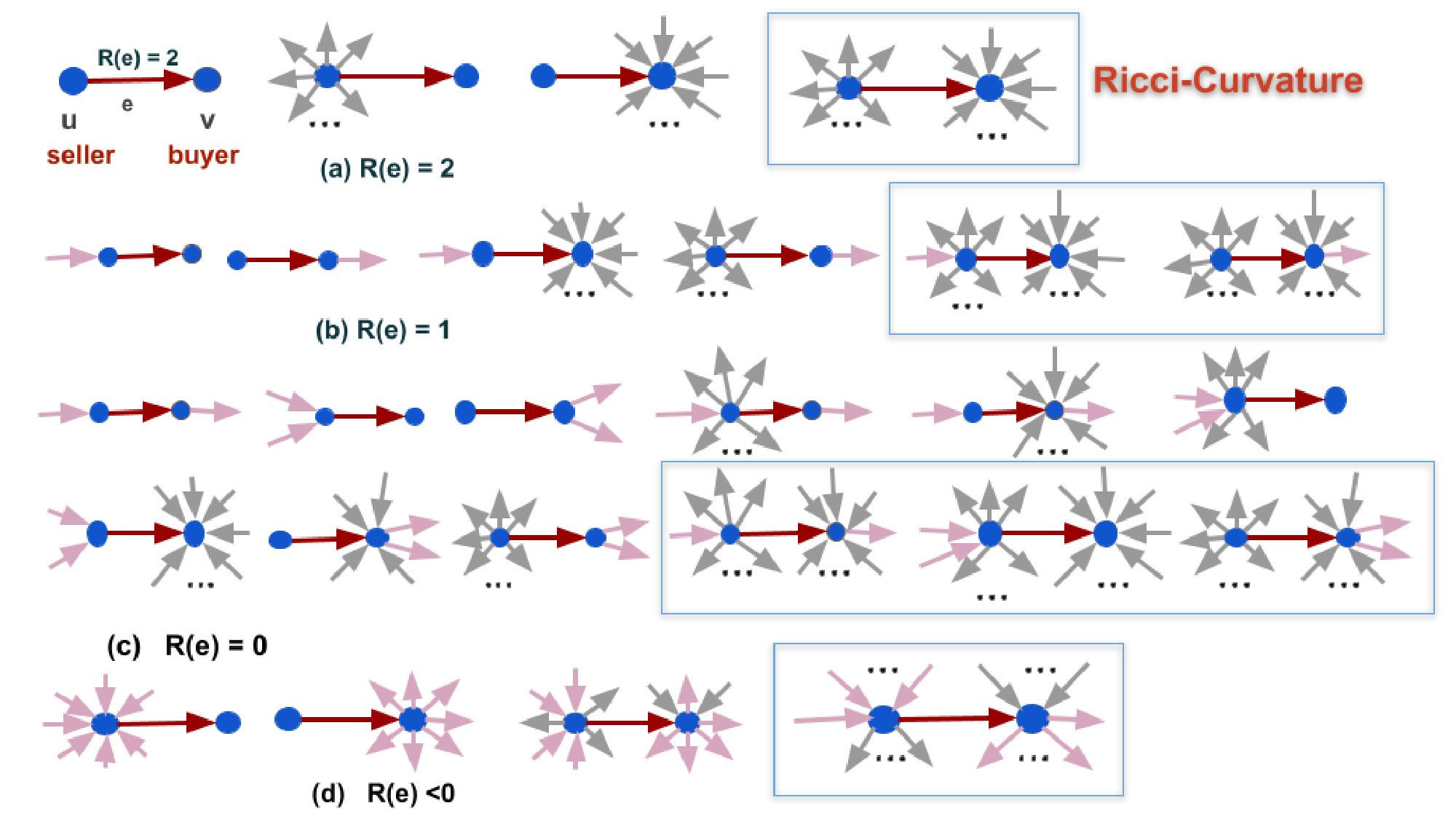}
\caption{From an economic perspective, we get the Forman-Ricci curvature ($R(e)$) of a transaction ($e$) between two wallets $u$ and $v$ to be (a) $R(e) = 2$ when the wallet $u$ can not buy any token and wallet $v$ can not sell any token; however, wallet $u$ can sell and wallet $v$ can buy tokens. (b) Similarly, $R(e) = 1$, we observe the same economic scenarios that we observe in $R(e) = 2$, with additionally, wallet $u$ can buy once or wallet $v$ can sell once. (c) In the case of edge with $R(e) = 0$, the wallet $u$ can buy at most twice, and wallet $v$ can not sell, or wallet $u$ cannot buy, and wallet $v$ can sell at most twice. However, the wallet $u$ can sell, and the $v$ can buy tokens. (d) Edge will have $R(e) << 0$ when wallet $u$ can buy and wallet $v$ can sell tokens. In other words,  $R(e) << 0$ when in-degree of $u$ and the out-degree of $v$ is high (Eq. (\ref{ricci_directed})). The edges in pink color contribute to the Forman-Ricci curvature of the edge $e$ (red) under consideration.}
\label{ricci_curvature_schematic}
\end{figure*}
In other words, the increase in the trading activity among small traders and among the big traders, and at the same time, a decrease in the trading activity between big traders and small traders has resulted in the co-movement of the ratios. From the above analysis of trading activity before and after the crypto crash, there is an evolution in the trading behavior of the traders. Before the crash, small traders perform most of the transactions with the big traders, but after the crash, small traders make most of the transactions among themselves. Also, there was an increase in trading activity among the big traders after the crash. Further from the dynamics of ratios, we observe a stronger co-movement during the pandemic period, which indicates the absence of a significant impact of COVID-19 on the trading behavior of the traders in the Ethereum platform. However, volatility exists in the evolution of the new ERC20 token inclusion to the platform during the COVID-19 period, whereas, after the crypto crash, the dynamics remained constant until July $2020$ (Fig. \ref{token_evolution}). Note that the key difference between correlation and regression is that correlation measures the degree of a relationship between two independent variables. In other words, the correlation between two variables captures how both variables are related. In contrast, regression is how one variable affects another. Both of the measures can not say whether variables are directly interacting with each other or not. 

\subsection{Forman-Ricci curvature analysis}
Now we use discrete Forman-Ricci curvature of networks introduced by R. Forman \cite{ricci_curvature_2018} to provide better insight into the trading behavior of the system. Forman-Ricci curvature is an edge-based concept that measures how fast edges spread in different directions \cite{ricci_curvature_2018}. Importantly, edges with negative curvature are vital in spreading information in a network. Previously, it has been used to characterize complex networks, which yield insights into their dynamical structure \cite{forman_ricci_dynamics_2017}. 
Since our networks are directed, we use the Forman-Ricci curvature of directed networks. The curvature of a directed edge $e$ of weight ${\omega_{e}}$, $u \rightsquigarrow v$ is defined as follows:
\begin{equation}\label{ricci_main}
R(e) = \omega_{e}\left( \frac{\omega_{u}}{\omega_{e}} - \sum_{e_{u} \sim{e}} \frac{\omega_{u}}{\sqrt{\omega_{e}\omega_{e_{u}}}} \right) + \omega_{e}\left( \frac{\omega_{v}}{\omega_{e}} - \sum_{e_{v}\sim{e}} \frac{\omega_{v}}{\sqrt{\omega_{e}\omega_{e_{v}}}} \right)
\end{equation}
\begin{figure}[tbh]
    \centering
\includegraphics[width=6in, height=4in]{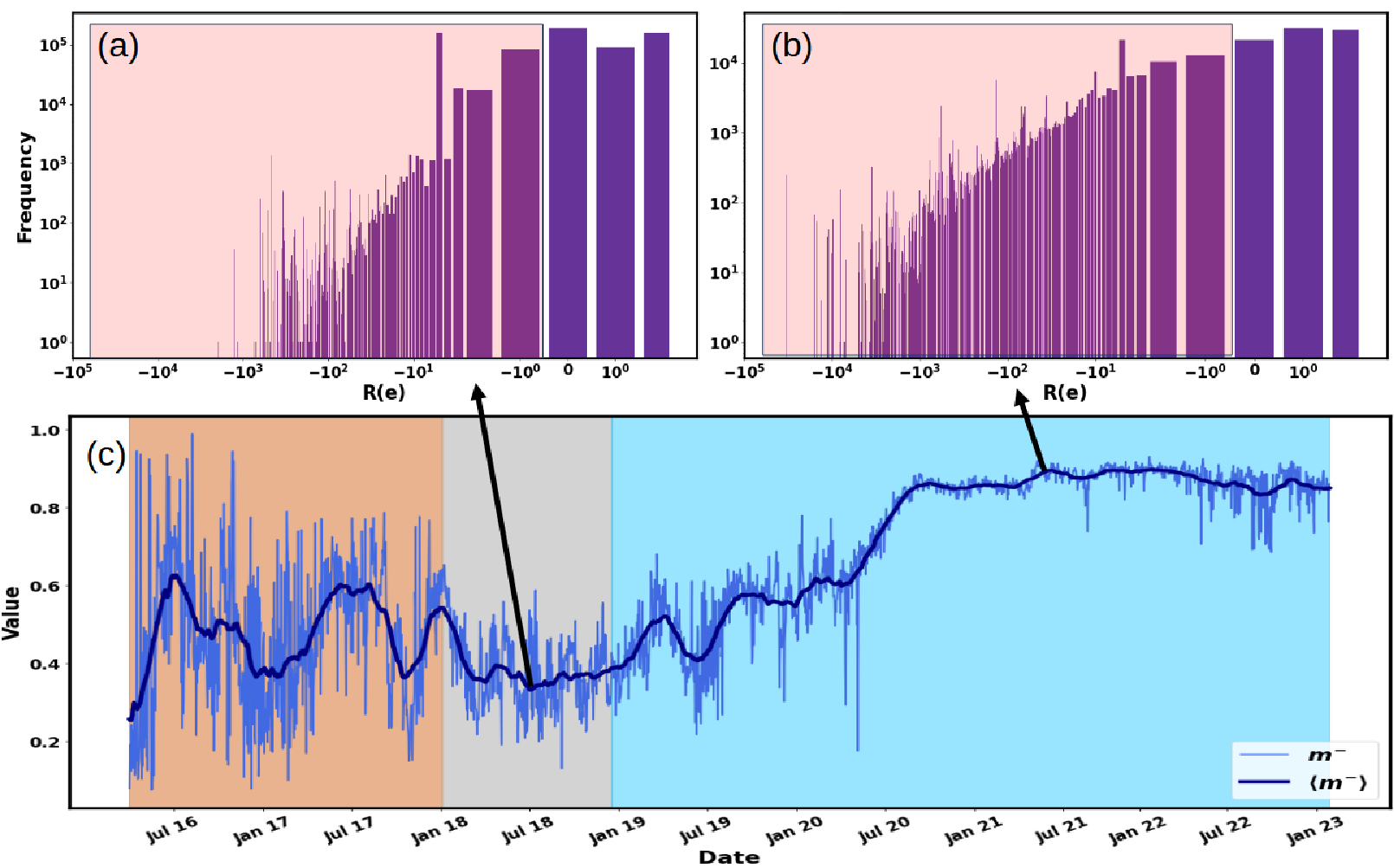}
\caption{Forman-Ricci Curvature of buyers' and sellers' behavior. We calculate the Forman-Ricci curvature ($R(e)$) of each edge ($e$) in a day using Eq. (\ref{ricci_directed}). We consider two snapshots of the $R(e)$ vs. frequency plot from July $2016$ to January $2023$. (a) the $R(e)$ vs. frequency plot for $12^{th}$ July $2018$ shows fewer spreads of the negative curvature values, where $n=336542$, and $E=818739$. On the other hand, (b) $12^{th}$ June $2021$ plots large spreads the negative Forman-Ricci curvature, where $n=254903$, $E=943758$. We represent the total number of edges $m_t=m_t^{+}+m_t^{-}$, where $m_t^{+}$ and $m_t^{-}$ are the number of edges with positive and negative Ricci curvature on day $t$ and after normalizing $m_t^{+}+m_t^{-}=1$. (c) the light blue line represents the fraction of negative Forman-Ricci curvature ($m^{-}$) contribution from daily transaction graphs. The dark blue line represents the moving average value ($\langle m^{-} \rangle$). We observe that the fraction keeps increasing after July $2018$ and becomes stable during the COVID-19 pandemic. The moving average window size is $70$ and calculated as in Fig. \ref{Rin_Rout}.}
\label{ricci_curvature_dynamics}
\end{figure}
where $e_{u}$, $e_{v}$ are the edges connected to node $u$, $v$ and $\omega_{e_{u}}$, $\omega_{e_{v}}$ are weights associated with the edges, respectively. Here, we only consider those directed edges that terminate at node $u$ and originate at node $v$. Since edges are unweighted, the above expression (Eq. (\ref{ricci_main})) reduces to
\begin{equation}\label{ricci_directed}
    R(e) = 2 -  \mathrm{indeg}(u) - \mathrm{outdeg}(v)
\end{equation}
where $u$ is the seller wallet, $v$ is the buyer wallet and $e$ is the transaction from $u$ to $v$. Here, $R(e) \leq 2$ as $\mathrm{indeg}(u)\geq 0$ and $\mathrm{outdeg}(u)\geq 0$. The curvature infers the structural properties of a network. Fig. \ref{ricci_curvature_schematic} shows some examples of the respective curvature of edges and the structure around them. 

The positive curvature of an edge $e$ infers limited types of trading activity between seller and buyer (Fig. \ref{ricci_curvature_schematic}). For instance, if a seller wants to buy more than $2$ times or a buyer wants to sell more than $2$ times, it can not be captured by the positive curvature (Fig. \ref{ricci_curvature_schematic}(a-c)). In other words, positive curvature refers to buyer-seller interaction with other traders in an isolated or restricted manner. There are few in-degree of the seller and fewer out-degree of the buyer, so wealth flows across the wallets in the network will be very slow and sometimes localized among peers (Fig. \ref{ricci_curvature_schematic}). On the other hand, the negative curvature of an edge $R(e) << 0$ refers to various trading activities carried out by the seller and buyer. It infers that the seller and the buyer can buy and sell multiple times. Therefore, increasing negative curvature infers dispersion of wealth across the network.

We calculate the fraction of edges {\bf ($m^{-}$)} contributing to the negative Ricci curvatures for the daily transaction network (Fig. \ref{ricci_curvature_dynamics}). We observe an increase in the fraction over time; it signifies an increase in the trading activity among the traders where simultaneously the seller sells and buys the tokens (Fig. \ref{ricci_curvature_dynamics}(c)). Similarly, buyers can also buy and sell the tokens. It shows the evolution in the behavior of the traders, where before the COVID-19 pandemic, most of the sellers only sold the tokens and buyers only bought the tokens, which resulted in a small percentage of edge with negative Ricci curvature, thus resulting in large positive Ricci curvature, and wealth localizes among the buyers (Fig. \ref{ricci_curvature_dynamics}(c)). However, after the crypto crash and during the COVID-19 pandemic, sellers and buyers both performed buying and selling activity which led to an increase in the percentage of edges having negative Ricci curvature value. Notably, one can observe that during both events, the number of daily transactions remains stable; only the trader's behaviors change.

\section{Conclusion}
In conclusion, the evolution of transaction graphs within the Ethereum blockchain's ERC20 token financial ecosystem from simple token transfers to complex DeFi protocols \cite{Decentralized Finance} reflects on the growth, complexity, and innovation occurring in the tokenized economy. Understanding and harnessing the insights from transaction graphs will be pivotal in addressing scalability challenges, fostering regulatory compliance, and unlocking opportunities for decentralized finance and digital asset utilization.

Using complex network analysis and differential geometry tools, we analyzed the dynamic evolution of transaction graphs in the ERC20 token financial ecosystem. We observed the evolution in the trading activity of the traders and the dynamics of ERC20 tokens in the financial ecosystem. We focused here on two big events - the $2018$ crypto crash and the COVID-19 pandemic. We started the investigation by analyzing the evolution of wallets, transactions, and tokens for the period of November 2015 to January 2023. There existed a constant addition of new tokens to the financial ecosystem until the pandemic; however, after that, there were fluctuations. Our analysis of the daily transaction graphs unveiled that before the crash, the trading activities of the traders led to the localization of wealth among individual traders. However, after the crash and during the pandemic, the change in trading activity by most traders led to the dispersal or continuous flow of wealth over the network. 

Here though, we used the extreme points of the degree distribution, incorporating other variables ($N_{\alpha,t}^{out}$ and $N_{\beta,t}^{in}$) in the analysis can provide more insight into the system which requires further investigation. Moreover, we use $4$  fields from the extracted data, and including other data, fields can provide greater insights into the financial ecosystem's underlying features. For instance, if we include the `value' field, the transaction networks become weighted and can provide insights into the flow of wealth in the financial ecosystem during the black swan events. Here, we examined the pairwise interactions of the transaction data, which cannot capture the higher-order interactions of the traders' behavior \cite{higher_order_interactions}. We intend to pursue higher-order interactions, which will provide insights into the existence and role of simultaneous many-body interactions in the financial market. As financial ecosystems continue to mature, further research and innovation in transaction graph analysis will be essential to unlock the full potential of ERC20 tokens and drive the adoption of decentralized applications built on the Ethereum blockchain.    


\section{Acknowledgement}
MP is thankful to the Complex Systems Lab (IIT Indore) members for the useful discussion. SJ acknowledges DST grant $SPF/2021/000136$. PP acknowledges SERB grant $TAR/2022/000657$.

\end{document}